\documentclass[12pt]{article}%
\usepackage{amsmath,latexsym}
\usepackage{graphicx}
\usepackage{amsmath}
\usepackage{amsfonts}
\usepackage{amssymb}%
\begin{document}
\title{{Determination of the mass and energy of primary cosmic rays above 100 TeV}}
\author{R. K. Dey*$^1$, S. Dam$^1$, S. Ray$^1$, A. Basak$^1$ and P. Chattopadhyay$^2$\\
\footnote{Corresponding Author E-mail: rkdey2007phy@rediffmail.com}
\small $^1$Department of Physics, University of North Bengal, Siliguri,\\ 
\small WB, INDIA 734 013\\
\small $^2$Department of Physics, Cooch Behar Panchanan Barma University, WB, INDIA}
\date{}
\maketitle
 
\begin{abstract}\noindent
This analysis aims to determine the mass composition and energy of cosmic rays at energies above 100 TeV based on the lateral distribution of extensive air showers. Here, we propose quite a few air shower observables for reconstructing the mass and energy of the primary particles. The present reconstruction uses a detailed Monte Carlo simulation for cosmic ray induced air showers in KASCADE and NBU types surface arrays of particle detectors. Some of the observables obtained from this analysis of simulated data are used to infer the nature of the primary particles from a comparison with KASCADE and/or NBU data. It is expected that the determination of primary energy of a cosmic-ray shower may deliver a better accuracy compared to standalone analysis using shower size or S600 or S500 or $N_{\rm {pe}}$ etc, owing to strong fluctuations in the EAS development. Moreover, the present study might be useful to discriminate between hadronic cosmic rays and primary gamma rays, and to measure the cosmic ray all-particle energy spectrum.\\                  

\end{abstract}

\section{Introduction}\label{S:Introduction}
The astrophysical models on the origin, acceleration and propagation of primary cosmic rays (PCRs) beyond $100$ TeV gain impetus constantly from two important sources of information. These are the energy spectrum and mass composition of PCRs obtained from various air shower experiments. Hence, the mass composition and energy spectrum of cosmic rays (CRs) are crucial for understanding the origin of the PCRs including their acceleration and propagation mechanisms. 

The basic strategy of the present analysis is to search some mass and energy sensitive air shower observables by exploiting different characteristics of extensive air showers (EASs) initiated by charged primaries \emph{e.g.} protons and irons, and  gamma rays, around the \emph{knee} region from a detailed Monte Carlo simulation. Air shower experiments, \emph{viz.} KASCADE \cite{rd1}, GRAPES-3 \cite{rd2}, ARGO-YBJ \cite{rd3} have (or are being) contributed tremendously to the current knowledge on the mass composition and energy spectrum in the TeV to PeV range \cite{rd4, rd25}. This was mainly done from the comparison between the measured properties of EASs initiated by the energetic particle and the detailed simulations of these particle cascades. This dependence on simulations is quite obvious and becomes a standard procedure to interpret air shower data, but it is also a hindrance. Alternatively, this dependence can be employed to test various aspects of high-energy hadronic interactions in present day accelerator energy ranges and even beyond. On the other hand, such a dependency introduces some limitations on how accurately the energy and mass of a primary particle can be reconstructed. Therefore, the more accurate the hadronic interaction models that are amended in simulation, the more accurate will be the estimate of the energy spectrum and mass composition.

Beyond $100$ TeV, the information about PCRs is achievable through the study of EAS which are cascades of mainly secondary electrons (${\rm e}^{+} + {\rm e}^{-}$), muons (${\mu}^{+} + {\mu}^{-}$) and hadrons (${\rm {h\bar{h}}}$) produced when PCR particles interact with atmospheric nuclei during their advancement towards the ground. In addition, air shower associated Cherenkov photons and fluorescence light are also produced simultaneously with the above components in an EAS. The ground based air-shower experiments equipped with scintillation detectors generally seed their measurements with two basic EAS data which are known to be densities and arrival times of electrons and/or muons. The crucial EAS parameters such as the shower size ($N_{\rm e}$), the EAS core ($x_{\rm o}$, $y_{\rm o}$) and the shower age ($s$), are obtained by applying the shower reconstruction to the electron density data using the Nishimura-Kamata-Greisen (NKG) type lateral density function (LDF) \cite{rd5}. The shower direction, i.e. the zenith angle ($\Theta$) and the azimuth angle ($\Phi$) are obtained using the shower front plane fitting with arrival time information of electrons from timing detectors. In the EAS experiments which are equipped with a {\lq hybrid\rq} detector system, EAS components such as muons ($\mu^{\pm}$), hadrons ($h\bar{h}$), air shower associated Cherenkov photons and arrival time of Cherenkov shower front, and fluorescence light are also measured. These measurements provide important EAS parameters such as the muon size ($N_{\mu}$), Cherenkov photon content ($N_{\text{Ch}}$) and the shower maximum ($X_{\text{max}}$), and also the longitudinal shower age ($s_{\parallel}$).

Some results of the present analysis include the \emph{knee} energy region which is of particular interest in astroparticle physics because one of the important changes in astrophysical processes is expected to occur, which would introduce a spectral break to the all-particle energy spectrum of PCRs \cite{rd4}. Present understanding takes into account the \emph{knee} to be an imprint of galactic CRs. The point of inflection at the \emph{knee} suffices to the maximum energy that can be reached at supernova remnants within our galaxy \cite{rd6}. The highest energies ever observed in the EAS experiments which are equipped with a {\lq hybrid\rq} detector system  (\emph{e.g.} Auger, Akeno, \emph{etc.}) could not have originated even from the most powerful supernova remnants. Hence, a transition from these sources to at least one additional population is obvious. The most acceptable proposition is that the ultra-high energy CRs ($E_{0} \geq 10^{5}$ TeV) are of extragalactic origin \cite{rd7}. However, a recent result by ARGO-YBJ experiment indicates the existence of a low-energy \emph{knee} at energies below 1 PeV from the measurement of primary p and helium spectrum \cite{rd25}. The discontinuity at source region around the \emph{knee} is believed to be linked with a $Z$ dependent variation in the contributions of the various elements/nuclei to the PCR energy spectrum. This might be due to either a concentration of individual element or some relative abundances of various elements \cite{rd8}. Hence, measurements of the mass and energy of PCRs around the \emph{knee} are very important for proper understanding of the origin of this spectral feature.

The central idea of this work is to use $N_{\rm e}$ and $N_{\mu}$ simultaneously as one of the single pair of observables for the estimation of the mass and energy of PCRs using simulated EAS data in the concerned energy range. A new parameter $\emph{f}$ is introduced that actually measures the ratio of electron densities taken within two suitable distance bands from the shower core. The possibilities of these EAS observables including the hadron size ($N_{\rm h}$, in addition to $\emph{f}$) as estimators of CR mass composition, will also be explored using Monte Carlo simulations.  In this context, the paper considers a few observed results (published results on mass composition) from two sea level experiments \emph{i.e.} NBU and KASCADE for the purpose of comparison. The NBU air shower array, located at North Bengal University campus, India (latitude $26^{\rm o} 42'$ N, longitude $88^{\rm o} 21'$ E, 150 m a.s.l., area 2000 m$^2$), was being operated during 1980 - 98 \cite{rd9}. The KASCADE experiment, located at Forschungszentrum Karlsruhe, Germany (latitude $49.1^{\rm o}$ N, longitude $8.4^{\rm o}$ E, 110 m a.s.l., area 40000 m$^2$), has stopped its operation in December 2015 \cite{rd1}.

This paper also indicates several distinguishing features between the gamma ray and normal CR nuclei initiated EASs in the TeV - PeV region with the help of some of the EAS observables which have already been introduced above. In astroparticle physics and gamma ray astronomy, the discrimination of gamma ray induced air showers from hadronic background is still an interesting research area in the TeV - PeV energy range and beyond in order to understand the origin of PCRs.

In this paper, Sec. \ref{S:mon} describes the important considerations adopted in the simulation. The shower reconstruction for EAS data analysis and the estimation of reconstructed average local electron density (ALED) are explained in Sec. \ref{S:ana}. Sec. \ref{S:dis} presents our results with discussion from the mass and the energy estimation procedures of individual primaries using suitable EAS observables. Finally our conclusions are given in Sec. \ref{S:con}.

\section{Air shower simulation}\label{S:mon}
The shower events are simulated by combining the high energy ($> 80$ GeV/n) hadronic interaction models QGSJet01.c \cite{rd10} and  EPOS 1.99 \cite{rd11}, separately with the low energy ($< 80$ GeV/n) hadronic interaction model GHEISHA version 2002d \cite{rd12} in the underlying structure of the CORSIKA code 6.970 \cite{rd13}. The EGS4 \cite{rd14} program library has been used for the simulation of the electromagnetic particles. The atmosphere \cite{rd15} takes planar approximation in the air shower simulation according to the US-standard atmospheric model that works for the zenith angle up to nearly $70 \deg$. We have restricted the maximum zenith angle for the PCRs to $45$ deg. 

The simulated events form a data library that contains about 0.1 million showers each for proton, iron and gamma ray at the NBU location in the primary energy ($E_{0}^{\text{SIM}}$) range $100$ TeV to $3000$ TeV using both QGSJet01.c and EPOS 1.99 models. In addition, we have generated 30,000 EAS events for each primary component proton, iron, and gamma ray and about 15,000 He events with the model QGSJet01.c in the primary energy range from $100$ TeV to $3\times 10^{4}$ TeV, following a power law with a spectral index of -2.7 below 3 PeV and -3.0 above the value at the geographical locations of KASCADE and NBU. 

In our simulation, we have allowed EAS particles (with proper particle's Id) to fall upon two artificial arrays that are almost identical/comparable to two sea level experiments namely NBU and KASCADE detector arrays for our simulation. For NBU array, we have taken sensitive detecting area as $\sim 2000$ m$^2$ with physical dimension 50 m $\times$ 40m for the simulation. About 42 scintillation detectors are distributed in the array with detector spacing $\sim 8$ m, and each having a surface area $\sim 0.25$ m$^2$ in the sensitive zone of the array. On the other hand, we have used a rectangular array of 252 detectors covering an area of 40000 m$^2$ (200 m $\times$ 200 m) for simulation at KASCADE location. In the array detector stations are equally spaced by $13$ m and each with an area of $\sim 1$ m$^2$. For estimating muon size, we have actually counted number of muons that pass through each of the detectors placed in the array with some threshold energy values (setting 0.23 GeV for KASCADE while 2.5 GeV for NBU simulations). We assume nearly $100\%$ detector efficiency so that the reconstructed muon sizes do not deviate much from true muon sizes. As a consequence, a muon size dependent energy estimation is expected to be more accurate. A trigger is provided in the simulation by estimating electron densities in 4 and 8 central triggering detectors respectively for NBU and KASCADE arrays. Our simulation uses a trigger that actually corresponds a state when central trigger detectors register at least 4 electrons per m$^2$.

A mixed composition (Mixture-I) is prepared from the generated showers taking $50 \%$ proton, $25 \%$ helium, $25 \%$ iron events for better understanding of EAS observational results in the TeV - PeV range. A second mixture (Mixture-II) of showers is also prepared, containing $40 \%$ proton, $40 \%$ iron and $20 \%$ gamma ray showers generated in the TeV - PeV region only to check whether any distinguishing features between hadronic and gamma ray showers reveal.

\section{Monte Carlo data analysis method}\label{S:ana}

The EAS events corresponding to cores landing within 25 m and 50 m respectively from the centers of NBU and KASCADE array are taken for analysis. These cores are evaluated by taking weight average of 8 higher values from electron density data of an event in case of NBU simulation while for KASCADE simulation about 16 such type of density data are used. Hence, for each event we have obtained $N_{\rm e}$, $N_{\mu}$, $N_{\text{h}}$ \emph{etc.} directly from the simulation, and a gross core location along with muon/electron density data at discrete detector locations from the so called weight averaged/grossed core. Then using shower reconstruction, we have finally obtained required EAS observables for the study and reconstructed lateral density distribution of electrons/muons \emph{etc.} 

The shower reconstruction involves the application of the NKG type LDF for obtaining $N_{\rm e}$ and $N_{\mu}$ parameters. Here, we simply fit simulated/observed data by means of a $\chi^{2}$-minimization routine using gradient search technique using the NKG LDF. Other EAS parameters \emph{e.g.} $N_{\text{h}}$, $\Theta$ and $\Phi$ are taken directly from the simulated showers. We have found that the NKG function represents the simulated data satisfactorily over the whole radial distance except at very small distances, where the simulated densities are found a little higher than that predicted by the NKG function. Hence, the simulated densities only in the radial interval $9-100$ m are used, and fitted showers with reduced $\chi^{2}$ less than $4$ are only accepted for results. For primary energies around $100$ TeV, the extension of a shower is normally spread over within $\sim 50$ m core distance whereas it exceeds $\sim 100$ m for energies around the knee \cite{rd26}. Fig. 1 presents the simulated radial densities of electrons obtained with the QGSJet model and their comparison with the reconstructed curves for 500 TeV proton and gamma ray initiated showers. The errors in estimating the shower size in the interval, $N_{\rm e}: 10^{3}{\textendash}10^{5}$ is obtained as $\pm 0.07N_{\rm e}$ for the QGSJet01.c model. We have applied a standard least square derivation procedure (gradient search method) to Monte Carlo showers with random Poisson errors that would ultimately led to error distribution in $N_{\rm e}$ and other parameters.

\begin{figure}[tbp]
\begin{center}
\includegraphics[width=0.5\textwidth]{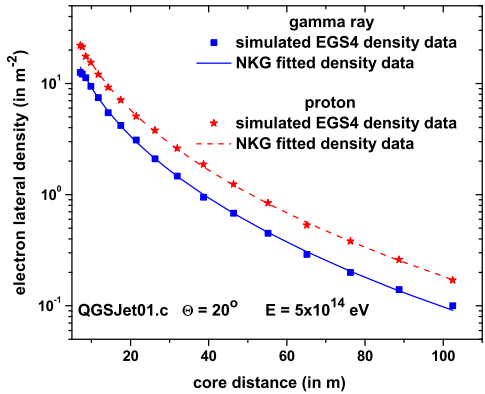}
\end{center} 
\caption{Comparison of the NKG fitted lateral density distribution of electrons with the simulated data for proton and gamma ray showers generated at $5\times 10^{14}$ eV energy.}
\end{figure}

Some earlier analyses of EAS data from various experiments indicated that the NKG or other LDFs could not give a good description of the radial distribution of electrons/muons with a single value of the age parameter at all distances \cite{rd16}. It was then suggested that LDFs could describe radial distribution of electrons arguably very well in short bands of radial distance with different suitable values of the age parameter. For implementing such an approach also, we first subdivide the whole radial distance $9{\textendash}100$ m on the shower plane into eight radial bands in which each one of the radial intervals contains five density data with 2.5 m separation each. Finally, we follow the above mentioned shower reconstruction using NKG type LDF for densities of electrons that are obtained within these individual small radial bands and after taking average over all directions. We have then obtained eight different mean values of the fitted electron density corresponding to these different radial bands for an EAS, and which we designate as the ALED.

\begin{figure}[tbp]
\begin{center}
\includegraphics[width=0.5\textwidth]{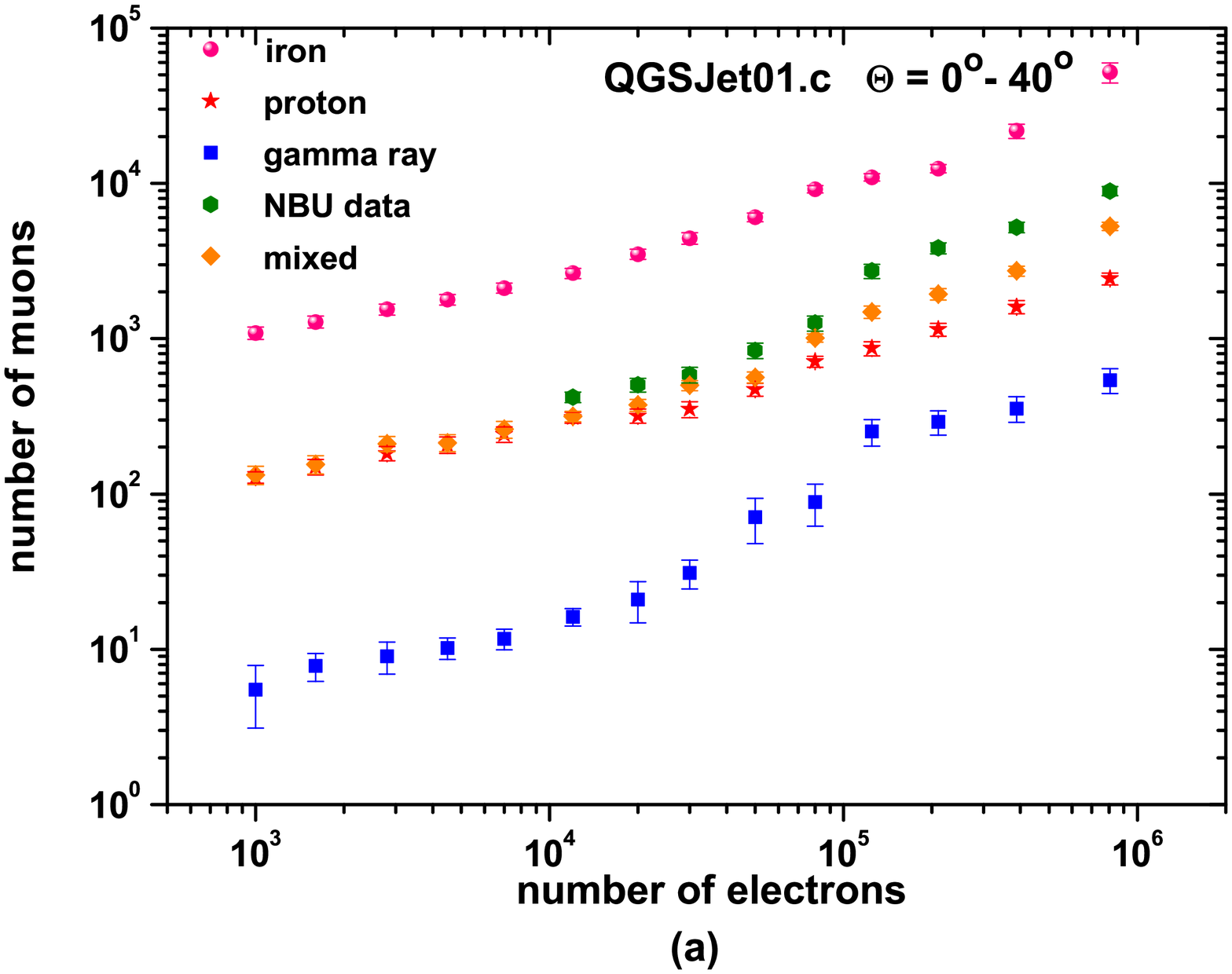} 
\includegraphics[width=0.5\textwidth]{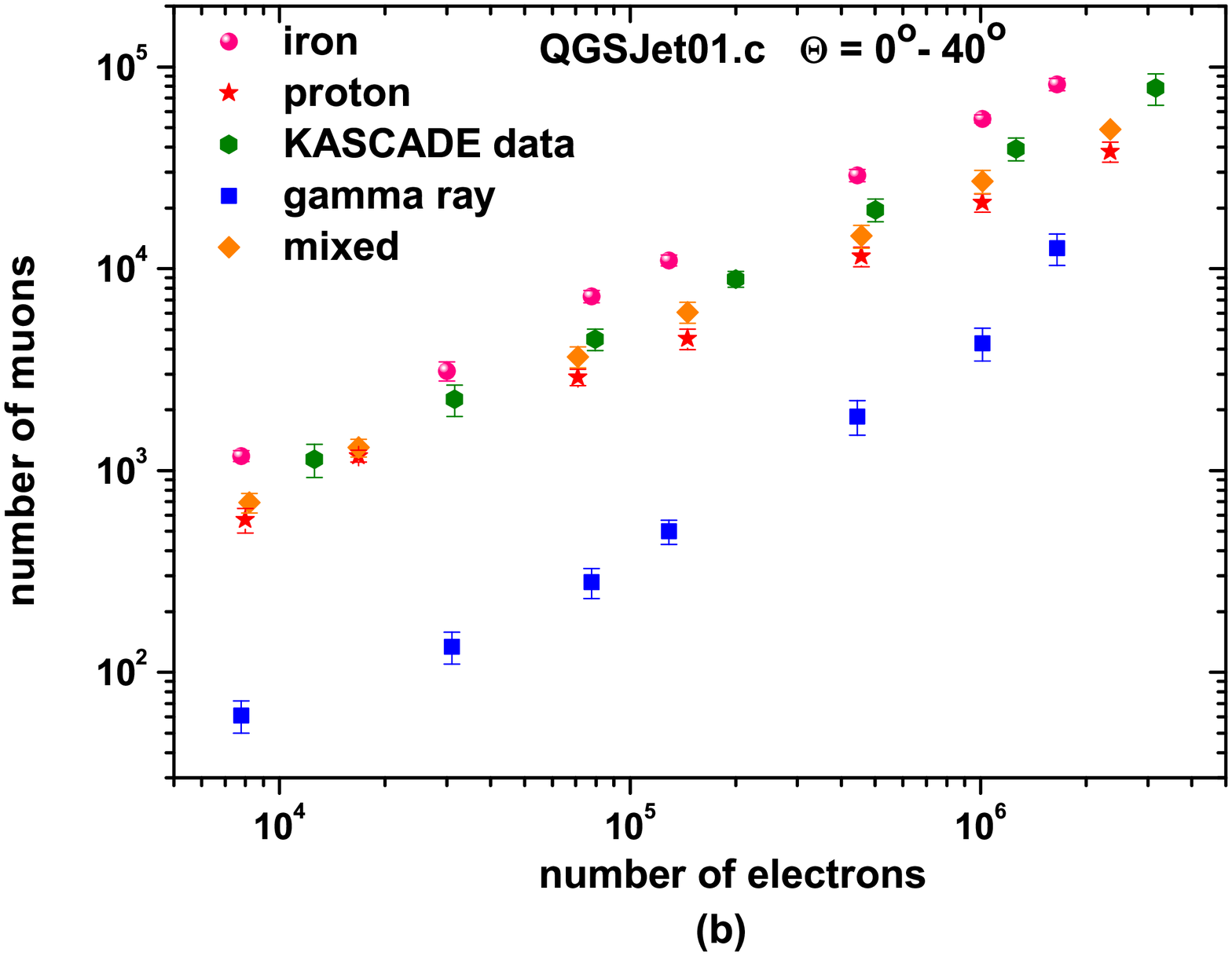} 
\includegraphics[width=0.5\textwidth]{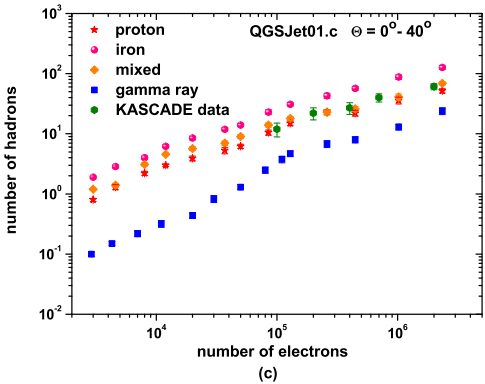} 
\end{center}
\caption{Mean number of muons and hadrons as a function of mean electron numbers for simulated proton, iron, mixed and gamma ray showers: (a) simulation and NBU data for muon-electron correlations; (b) same as Fig. (a) but at KASCADE level. In (c), simulation and KASCADE data for hadron-electron correlations.}
\end{figure}

\section{Results and discussion}\label{S:dis}

We have analyzed simulated EAS events generated at NBU and KASCADE locations for obtaining a set of primary mass and energy sensitive observables like $N_{\rm e}$, $N_{\mu}$, $N_{\mu}^{\text{Max}}$, $N_{\text{h}}$, $\emph{f}$ \emph{etc}. Our simulated results are then used for comparison with available  published results of NBU \cite{rd17} and KASCADE \cite{rd18} experiments.

\subsection{Electron-muon and electron-hadron correlations}

The most important indication of a gamma ray initiated shower is the presence of relatively less number of muons in the shower \cite{rd19}. The muon content can also be adjudged as a good distinguishing parameter among various CR/hadronic primaries. The production of electron-positron pairs are predominant when an EAS is initiated by a primary gamma ray. On the other hand, the production of muon pairs is inhibited by more than four order of magnitude due to larger muon mass. The only process is via the photo-production of hadrons where muons can occur to an appreciable amount in EAS initiated by gamma ray. 

Secondary hadron content of EAS is also useful to extract information on the mass composition of CRs. The mean number of muons and secondary hadrons as a function of the number of electrons are depicted in Figures 2a, 2b and 2c. It could clearly be recognized that curves for hadronic primaries are well above from that for the gamma ray in all the figures. It implies that the electron-muon and electron-hadron correlations could discriminate primaries; more efficiently the gamma ray showers from the hadronic showers. It should be however mentioned that for EAS experiments operating in the high energy region, the selection of primary gamma rays (typically, $\geq 100$ TeV or so) from a huge hadronic background is being made very often by measuring the muon content of showers, and is recognized as a promising technique for the purpose till date \cite{rd20}. However, such a technique becomes inefficient in regimes where the expected muon number for protons/nuclei initiated showers is low due to lower primary energy and/or with wider geometrical spread in the shower profile. In such a situation protons/nuclei initiated showers are eliminated from gamma ray showers through the estimation of EAS events which are \emph{secondary hadron-rich} with respect to the number of hadrons expected from gamma ray initiated showers \cite{rd21}.

In Figures 2a and 2b, we plot the electron-muon correlations from the simulations for gamma ray, proton, iron and mixed primaries at NBU and KASCADE levels. The corresponding observational results of these experiments are also used for comparison in those figures. The figures indicate that both the NBU and KASCADE data hint for a mixed composition and at the same time exhibit a gradual transition from light to heavy mass composition around the knee. The KASCADE hadron calorimeter could give a measure on secondary hadron size associated with an EAS. In Figure 2c, we have shown electron-hadron correlations for simulated and observed data for KASCADE to get a better idea on the primary CR mass composition based on hadrons. The data in the figure however do not exhibit any distinct signature on mass composition but favor a mixed composition across the knee.

These figures show that for NBU simulation, the \emph{knee} point corresponds the $N_{\rm e}$ and $N_{\mu}$ values close to $\sim 1.5\times 10^5$ and $\sim 10^3$ for proton initiated showers. For Fe showers, these values are $N_{\rm e} \sim 8\times 10^4$ and $N_{\mu}\sim 1.5\times 10^3$. These parameters take values as $1.6\times 10^5$, $2.5\times 10^4$ for p showers while for Fe, these are $5\times 10^4$ and $3.2\times 10^4$ respectively for KASCADE data.

For Figure 2a, the energy range used was $10^{14}-7.5\times 10^{15}$ eV. In Figures 2b and 2c, the energy range is taken as $10^{14}-2.5\times 10^{16}$ eV. The corresponding electron size lies within $\sim 7.5\times 10^{3}-3.1\times 10^{6}$ for KASCADE simulation while for NBU simulation, it follows a size range as; $\sim 10^{3}-8.25\times 10^{5}$.

\begin{figure}[tbp]
\begin{center}
\includegraphics[width=0.5\textwidth]{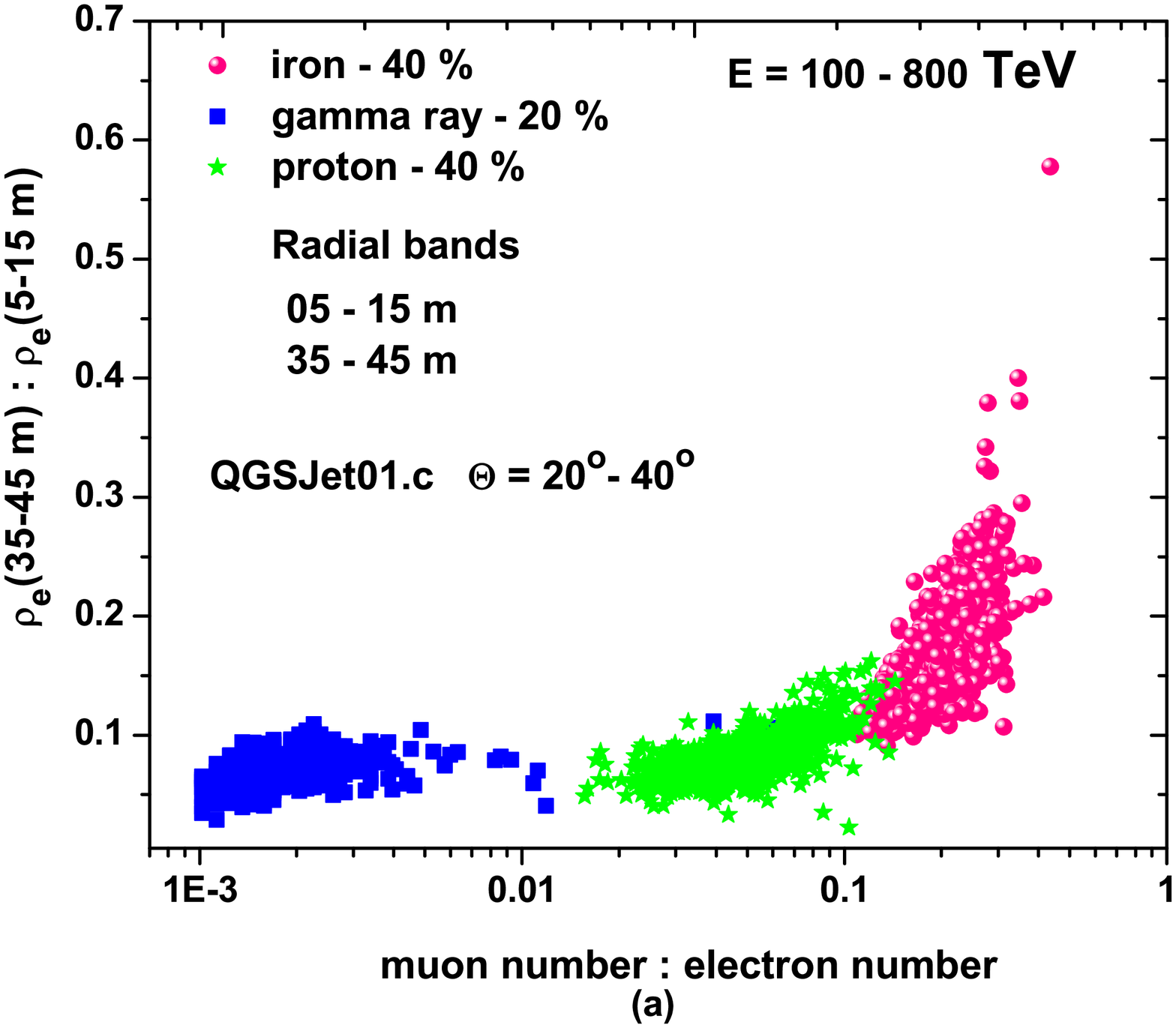} 
\includegraphics[width=0.5\textwidth]{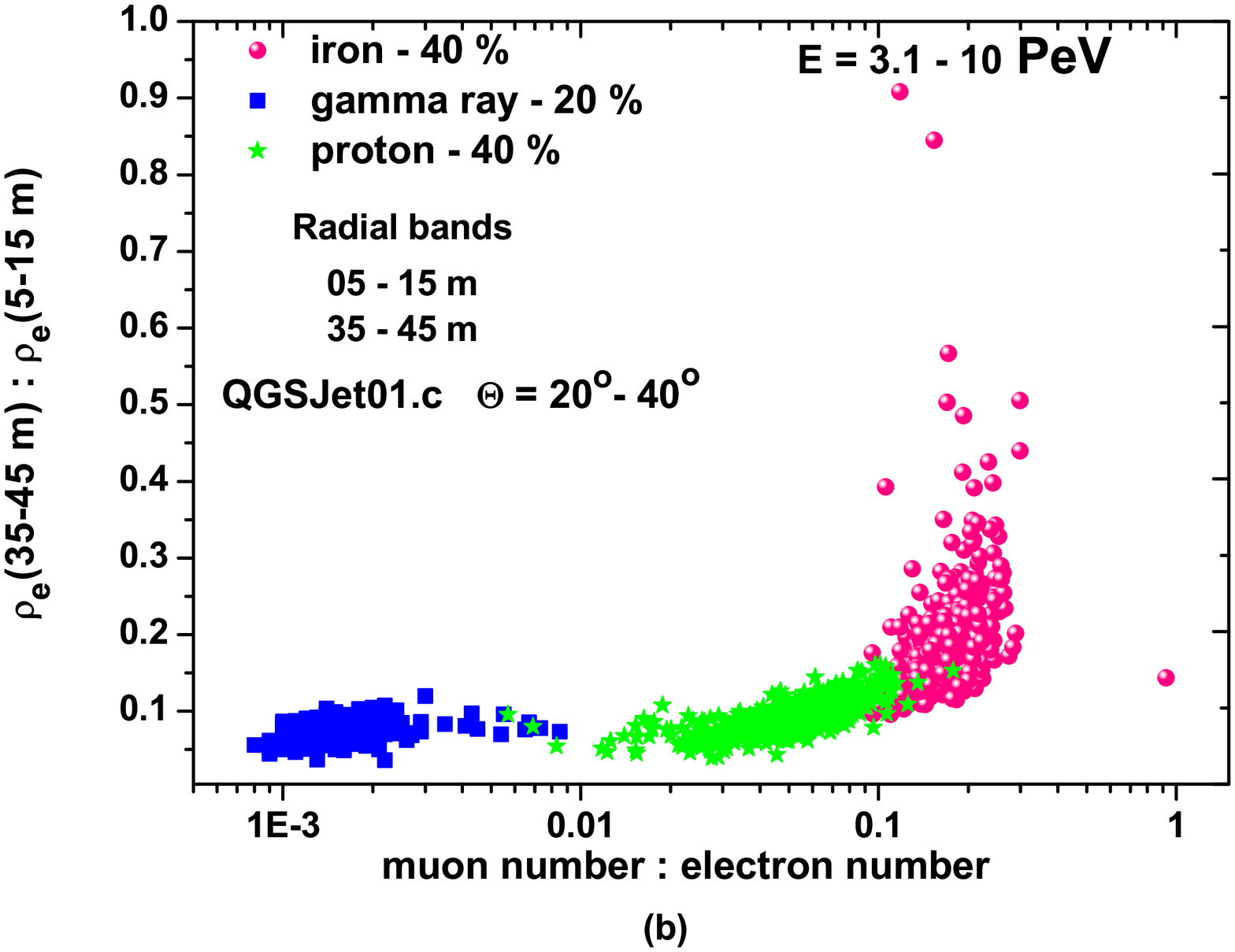}  
\end{center}
\caption{Variation of the ratio between the two electron densities estimated at two distance bands $5-15$ m and $35-45$ m with the muon number to electron number ratio at KASCADE level: (a) 100 - 800 TeV; (b) 3.1 - 10 PeV}
\end{figure}

\subsection{Variation of electron density ratio with muon to electron number ratio}

The lateral distribution of shower electrons at any observational level initiated by PCRs can be used to determine the mass of individual species. For the purpose, we have selected simulated showers those are found within a particular primary energy/shower size range corresponding to a particular $\Theta$-range from our simulated data library. From the LDF of each simulated EAS, we have estimated electron densities at five adjacent radial points in each of the two arbitrarily chosen core distance bands $5-15$ m and $35-45$ m using shower reconstruction to obtain finally two average ALEDs ($\rho_{1}$ and $\rho_{2}$) from the EAS core.

Number of muons produced by gamma ray showers is generically very less and therefore $N_{\mu}$ should not be treated alone as a statistically significant parameter as far as the gamma/hadron separation is concerned. In this situation, a multi-parameter approach using $N_{\rm e}$, $N_{\mu}$ and $\frac {\rho_{2}}{\rho_{1}}$ parameters is expected to be more effective to infer anything related to PCRs. Hence, any conclusion that can be drawn on gamma/hadron discrimination might be more reliable than the prediction from $N_{\rm e}$-$N_{\mu}$ correlation.    

We have plotted the ratio of two ALEDs ($\frac {\rho_{2}}{\rho_{1}}$) versus muon to electron number ratio ($\frac {N_{\mu}}{N_{\text {e}}}$) at energies below and above the \emph{knee} through Figures 3a and 3b for a mixed sample that contains $40 \%$ proton, $40 \%$ iron, and $20 \%$ gamma ray showers. It clearly suggests that these parameters and their mutual variations are quite useful to determine the mass of PCRs.  

The relative percentage of $\gamma$-ray flux to the proton flux was reported to be of the order of $\sim 10^{-5}$ at $10^{14}$ eV. Some other works predicted the ratio as: $\gamma/{\rm p}\sim 10^{-2}$ at $10^{16}$ eV \cite{rd27}. During their propagation in interstellar space, TeV - PeV gamma rays are being attenuated by the radiation fields (starlight, dust emission, cosmic microwave radiation \emph{etc}.) \cite{rd28}. This might be a possible reason for the poor percentage of primary $\gamma$-ray flux that could have been observed on earth. In comparison with the above fraction of gamma rays, Mixture-II ($40\%$ protons, $40\%$ iron and $20\%$ gamma rays) though appeared fully unrealistic but it can be used to assess the discriminating power of our adopted {\lq multi-parameter approach\rq}. 

\subsection{Primary energy reconstruction from lateral and longitudinal shower profiles}

The reconstruction of the energy of the PCRs is very crucial because the parameter has a direct connection with the origin, as well as acceleration and propagation mechanisms of PCRs. Different procedures have been adopted in shower data analysis by different experiments to estimate the PCR energy. For example, the estimation of the PCR energy for KASCADE-Grande is based on the estimation of a pair of observables, namely, the total charged particle number ($N_{\text{ch}}$) and the $N_{\mu}$ \cite{rd8}. The AGASA experiment adopted S600, which is the electron/charged-particle density at a lateral distance of 600 m from the core, as an energy estimator. An accurate estimation of the PCR energy requires a more accurate prediction from the simulation which is being used to interpret EAS data.

Among the several observables introduced in the present work, the $N_{\mu}$ is found reasonably a good primary energy estimator in the concerned energy region. From the longitudinal data sub-block of CORSIKA output for proton, gamma ray and iron simulated showers, it is noticed that the number of total muons remains nearly the same in an EAS after reaching the shower maximum. However, for gamma ray induced showers, attenuation of muons after the shower maximum becomes little faster than the case for protons or nuclei induced showers. These attenuation features of showers irrespective of their primaries lead to consider the $N_{\mu}$ as a primary energy estimator at any observable level beyond the point of shower size or muon size maximum. As $N_{\mu}$ is considered to be an energy estimator here, therefore, a more reliable and refined high-energy hadronic model has to be selected in simulations. The EPOS 1.99 model is used in our simulations because it involves an additional particle production source which induces a larger number of muons from simulations \cite{rd23}.

Although $N_{\mu}$ suffers very small attenuation due to atmospheric overburden mainly for inclined showers from the point of muon size maximum but this feature could serve our purpose for estimating the PCR energy using muon number at the observation level. A particular $N_{\mu}$ attenuates with $\Theta$ for the same $N_{\rm e}$ or $E$ due to the variation in the threshold energy of detected muons as $\propto sec~{\Theta}$ and the increase of the probability of muon decay with $\Theta$. This behavior of $N_{\mu}$ attenuation is basically incorporated to the so called constant $N_{\rm e}\textendash N_{\mu}$ method applied to data in many experiments. 

\begin{figure}[tbp]
\begin{center}
\includegraphics[width=0.45\textwidth]{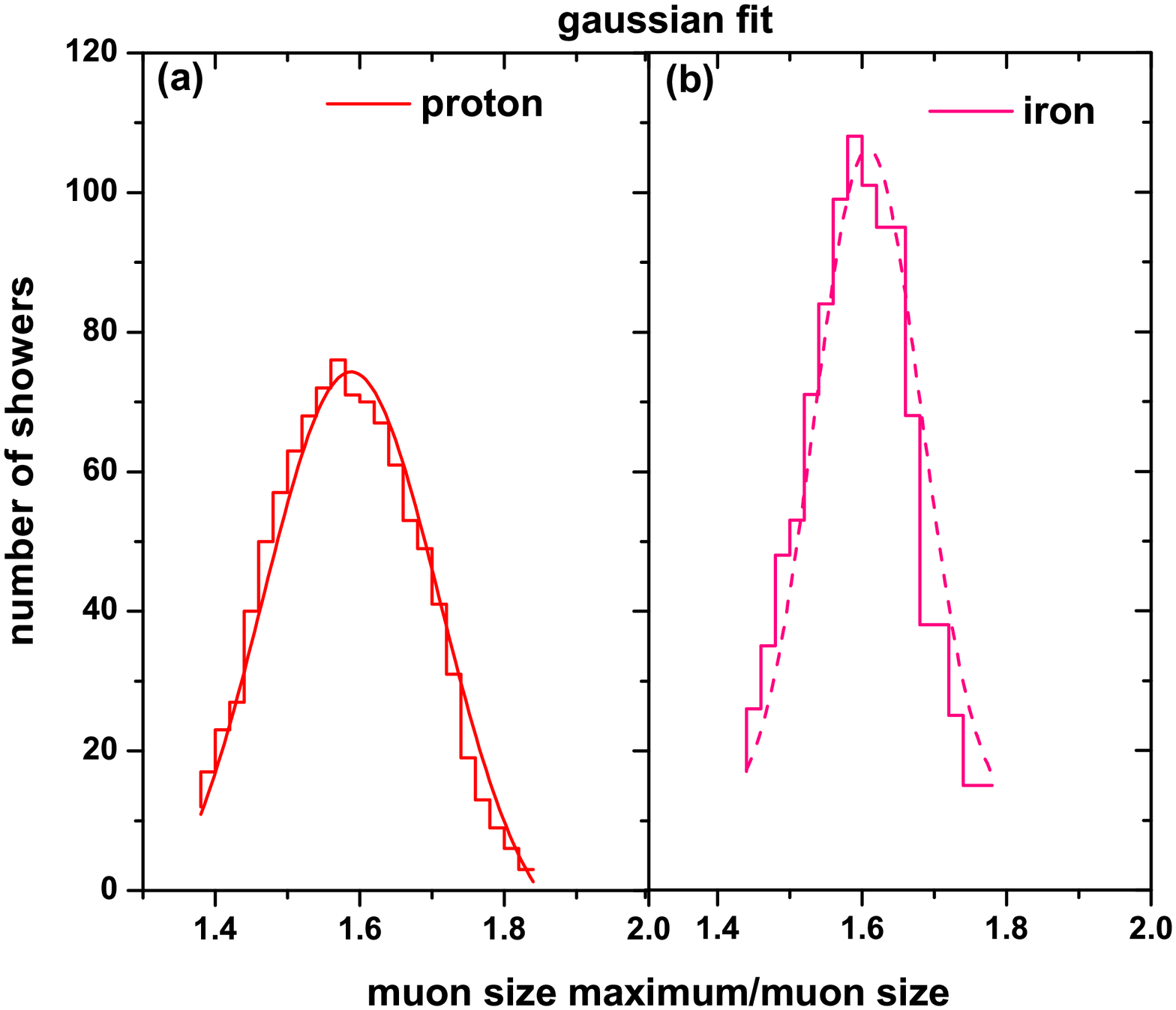} 
\includegraphics[width=0.45\textwidth]{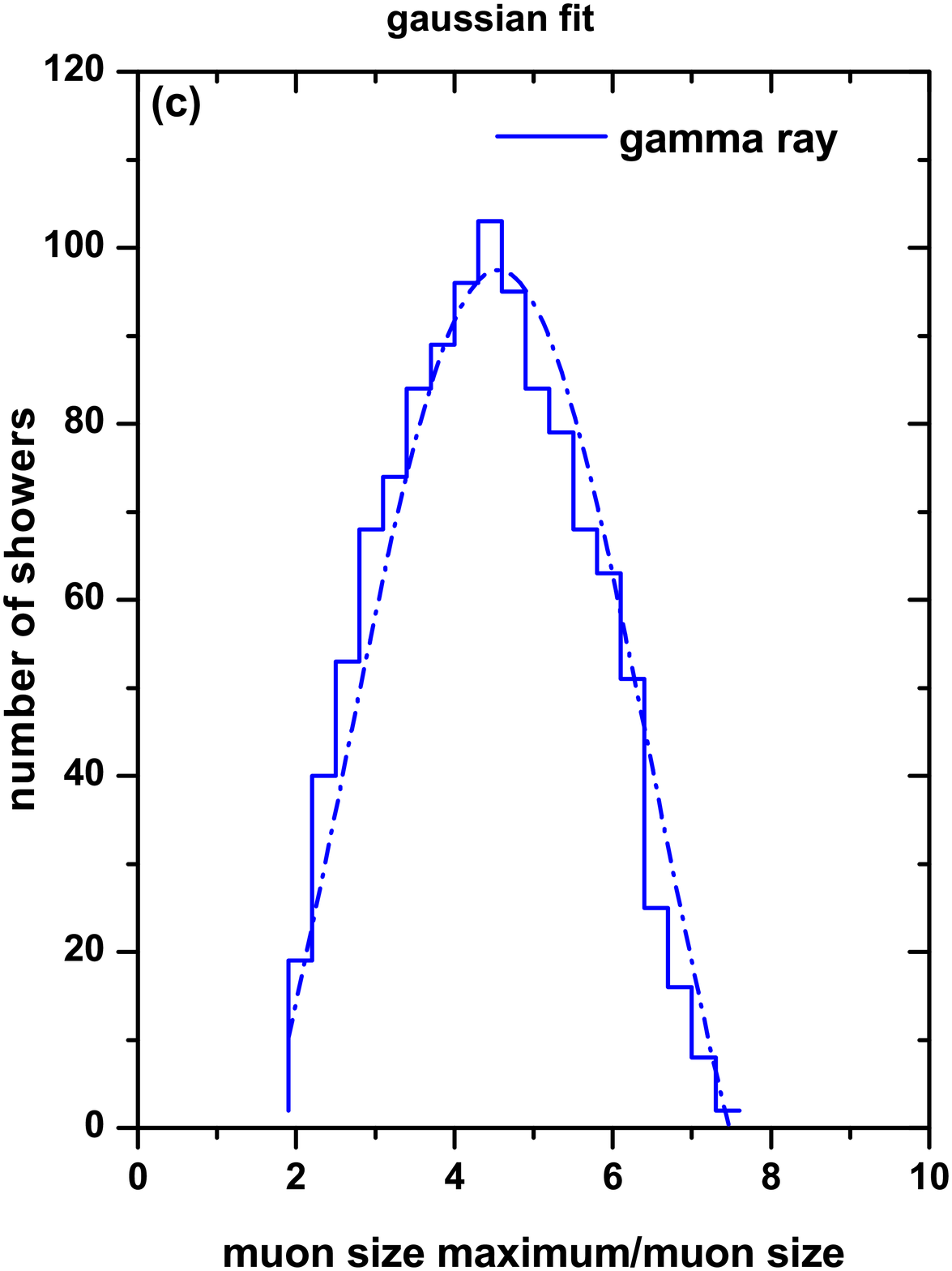} 
\end{center}
\caption{Frequency distributions of $\frac {N_{\mu}^{\text{Max}}}{N_{\mu}}$; (a) for proton; (b) iron; (c) gamma ray showers. We have considered showers in the primary energy range 200 - 800 TeV with $\Theta = 20^{\rm o}- 40^{\rm o}$. Models used are EPOS 1.99 and GHEISHA.} 
\end{figure}

The PCR energy estimation in the present analysis ultimately requires quite a few EAS observables from a ground-based EAS experiment \emph{viz.} $N_{\mu}$, $\Theta$ and $N_{\rm e}$. The $N_{\mu}^{\text{Max}}$ parameter however can only be obtained from the simulation employed in an EAS experiment. We have selected proton, iron and gamma ray initiated showers from our simulated shower library with some specific zenith angle and primary energy ranges, $\Theta = 20^{\rm o} \textendash 40^{\rm o}$ and $E = 200\textendash 800$ TeV here. The possible $\Theta$ range selection in the work provides quite a reasonable statistics of events from the whole range, $0^{\rm o} \textendash 45^{\rm o}$ in the simulation. Detection of gamma rays in the multi-TeV range provides an opportunity to know the possible origin of PCRs. In this multi-TeV scale the percentage of gamma ray flux is extremely low, of the order of $0.001 \%$ or less. Hence, the PCR energy reconstruction in the range of $200\textendash 800$ TeV might be useful in the area of multi-TeV gamma ray astronomy, and astroparticle physics.

\begin{figure}[tbp]
\begin{center}
\includegraphics[width=0.5\textwidth]{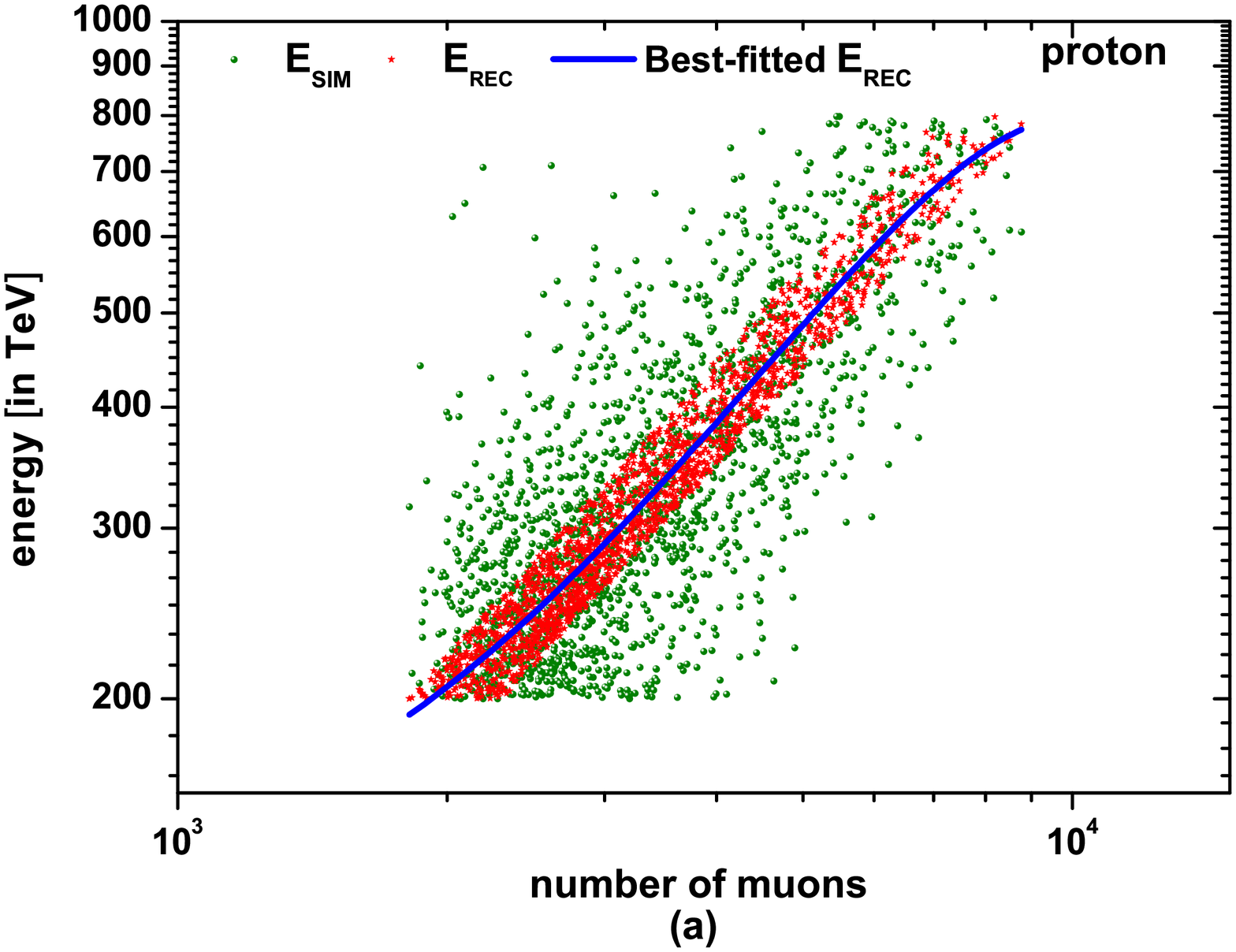} 
\includegraphics[width=0.5\textwidth]{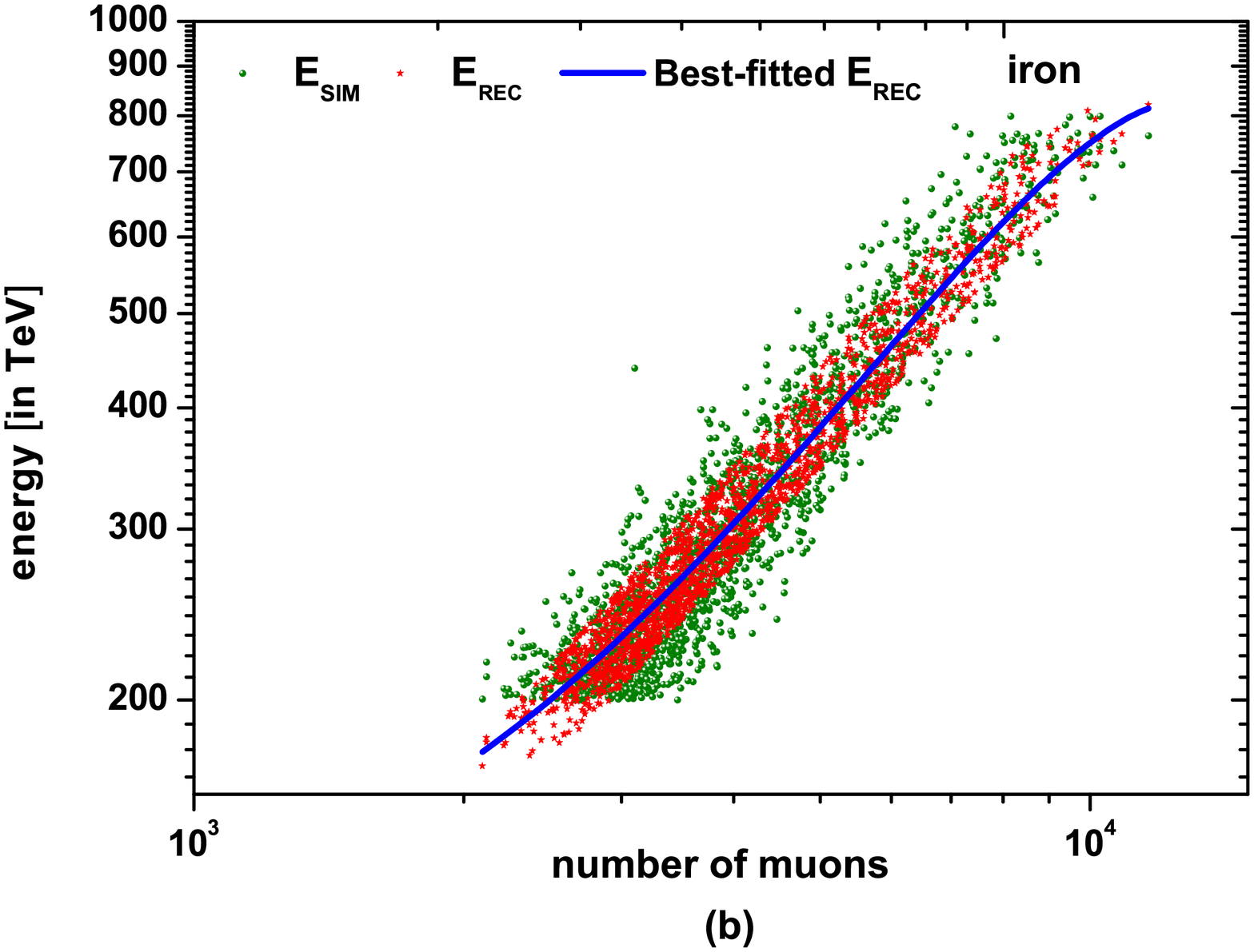} 
\includegraphics[width=0.5\textwidth]{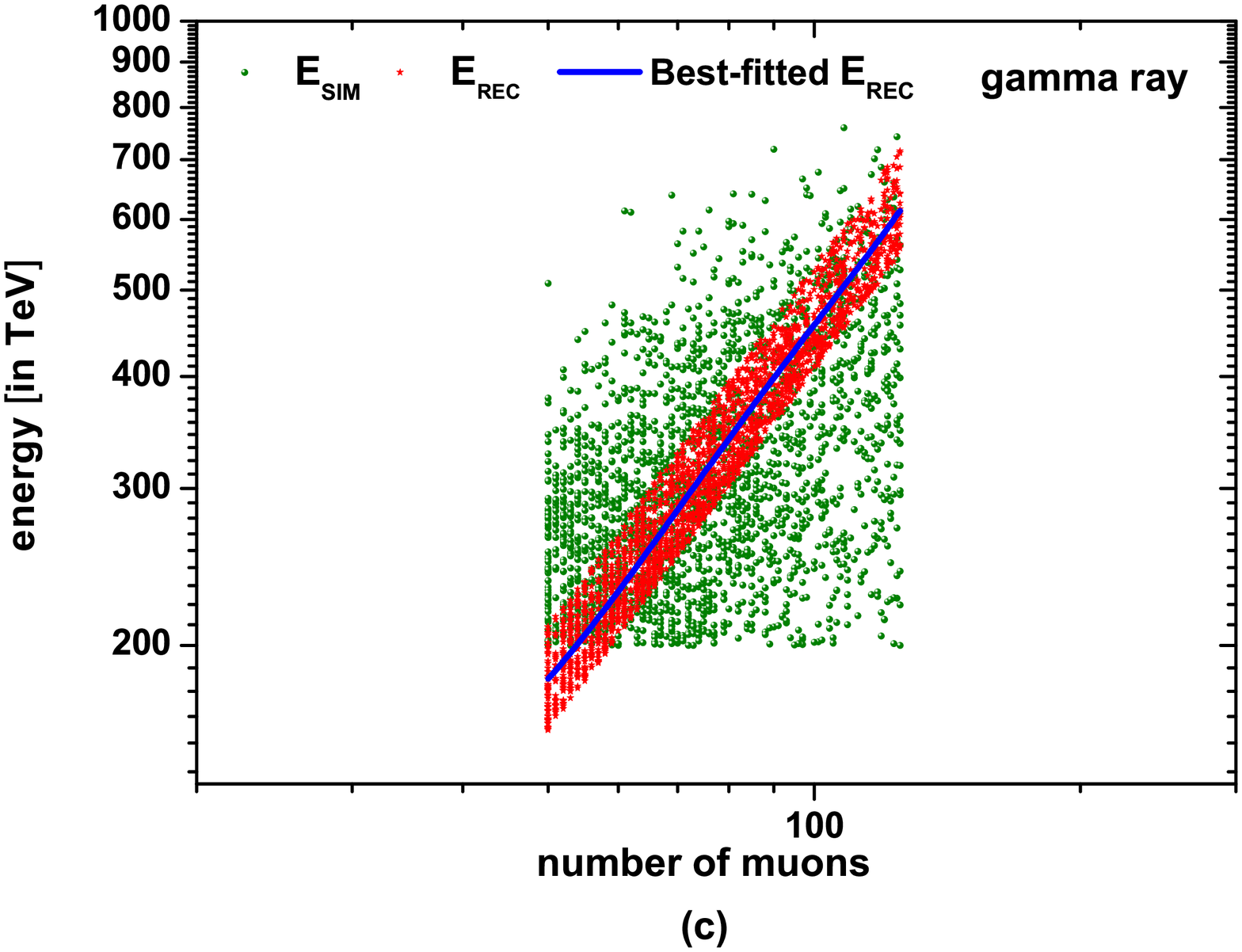} 
\end{center}
\caption{Correlations between the simulated primary energy and the produced muon number for (a) proton, (b) iron and (c) gamma ray induced showers. The red patch in each figure shows the reconstructed energy obtained from the energy parametrization according to the 
equation (3). The solid blue curves show the variation of the primary energy with number of muons according to the best-fit with polynomial functions in the $200\textendash 800$ TeV energy range. Models used are EPOS 1.99 and GHEISHA.} 
\end{figure}

We will now discuss how the information on $N_{\mu}$ or $N_{\mu}^{\text{Max}}$ and $\Theta$ can be utilized to obtain the PCR energy ($E_{\rm o}$). The available muon data in the longitudinal sub-block of CORSIKA for simulated showers give the variation of $N_{\mu}$ with atmospheric depth, which in turn provides the $N_{\mu}^{\text{Max}}$. Moreover, the important correlation of $N_{\mu}$ on the slant depth ($sec~\Theta$) is also necessary to determine $E_{\rm o}$. 

The $N_{\mu}$ attenuates with $\Theta$ for the same primary energy. Showers at constant energy could be used to obtain the $\Theta$-dependence of $N_{\mu}(\Theta)$. The procedure is applied through three steps. First, a particular $N_{\rm e}$ interval is selected (say, between $N_{\rm e} - 0.25N_{\rm e}$ and $N_{\rm e} + 0.25N_{\rm e}$). Secondly, frequency distributions of $N_{\mu}$ are made for different smaller $\Theta$ intervals. Finally, a superimposed region from the decaying part or tail of these various distribution curves or histograms of $N_{\mu}$ is set as $N_{\mu}$ range within which $N_{\mu}$ attenuation should be studied. Under this condition, the attenuation length ($\Lambda_{\mu}$) of muons in air in the selected $N_{\mu}$ range can be expressed by the following  
\begin{equation}
~R({\Theta})~=~R(0)e^{-\frac{X_{\mathrm o}}{\Lambda_{\mu}}(sec~{\Theta}-1)},
\end{equation} 
where $R(\Theta)$ and $R(0)$ are the rates of shower events in an arbitrary unit (a.u.) at a particular $\Theta$ and $\Theta = 0^{\rm o}$. The vertical atmospheric depth for the observed level is denoted by $X_{\text{o}}$. 

First, we will proceed for correlations between reconstructed energy ($E_{0}^{\text{REC}}$) and the prime energy estimator i.e. the $N_{\mu}^{\text{Max}}$. Later, by introducing a size-dependent scale factor, say $\beta$, the $N_{\mu}^{\text{Max}}$ is converted into $N_{\mu}$ in the energy reconstruction formula of $E_{0}^{\text{REC}}$. When attenuation characteristics of muons in air are considered then $E_{0}^{\text{REC}}$ can be expressed by the formula as follows,

\begin{equation}
~E_{0}^{\text{REC}}~=~\delta_{E}A(\Theta)(N_{\mu}^{\text{Max}})^{\beta},
\end{equation}
where $\delta_{E}$ is the overall energy conversion factor of $N_{\mu}^{\text{Max}}$ obtained from simulation. The $\Theta$-dependent scale factor $\text{A}(\Theta)$ appeared due to attenuation of $N_{\mu}$ for showers with $\Theta \neq 0^{\rm o}$. The factor $\text{A}(\Theta)$ has been found to be proportional to $e^{\frac{X_{\mathrm o}}{\Lambda_{\mu}}(sec~{\Theta}-1)}$, where $\Lambda_{\mu}$ takes different values due to in-congruent development processes of cascades induced by different primaries.

The $N_{\rm e}$ range needed for estimating $\Lambda_{\mu}$ from generated showers with $200\textendash 800$ TeV energy values is selected as $2\times 10^{3}\textendash 4\times 10^{3}$. The corresponding selected $N_{\mu}$ bins for proton, iron and gamma ray showers are $1.0\times 10^{3}\textendash 1.6\times 10^{3}$, $2.8\times 10^{3}\textendash 4.0\times 10^{3}$ and $75\textendash 100$ respectively. We have then estimated mean values of the ratio $\mathrm{\it{g}} = \langle\frac {N_{\mu}^{\text{Max}}}{N_{\mu}}\rangle$ taking into account all the generated events for proton, iron and gamma ray primaries separately. We have obtained these values for $\mathrm{\it{g}}$ close to $1.59$, $1.61$ and $4.55$ from the frequency distribution plots shown in Figure 4 for proton, iron and gamma ray showers.

We substitute $\delta_{E}$ (after absorbing $\it{g}$ into it) and also $\beta$, and $\rm A(\Theta)$ obtained from simulation into the eqn. (2). Finally, the relations between $E_{0}^{\text{REC}}$ (in TeV) and the pair $N_{\mu}$ and $sec~\Theta$ for proton, iron and gamma ray primaries are as follows:

\begin{equation}
~E_{0}^{\text{REC}}~=~\mathrm{a}(N_{\mu})^{\mathrm{b}}e^{\mathrm{c}(sec~{\Theta}-1)}
\end{equation}  

The values of $\Lambda_{\mu}$ appear in $\text{A}(\Theta)$ due to attenuation of muons in air are borrowed from simulation results of KASCADE using the constant $N_{\rm e}\textendash N_{\mu}$ method. These values are $982$, $1028$ and $901$ gm-cm$^{-2}$, with an average error of nearly $\pm{91}$ gm-cm$^{-2}$ that resulted from intrinsic fluctuations present in EAS development \cite{rd22}. The set ($\text{a},\text{b},\text{c}$) takes various values, as obtained from simulation, and those are; ($0.076,1.01,1.04$), ($0.053,1.026,0.97$) and ($0.756,1.33,1.136$) for proton, iron and gamma ray primaries in the concerned energy region.     

The above parametrization obtained for the PCR energy reconstruction from simulations showing slight differences from one type to another type of primaries only in terms of various factors. This nature is inherent because showers initiated by proton/nuclei undergo different developmental processes relative to gamma ray showers. On the other hand, a very little departure in the parametrization of iron from proton may be accounted by their higher developmental tendency in air. Iron showers have about $1.5$ times more muons than proton showers at the same energy \cite{rd24} which is a generic feature of EASs.

\begin{figure}[tbp]
\begin{center}
\includegraphics[width=0.5\textwidth]{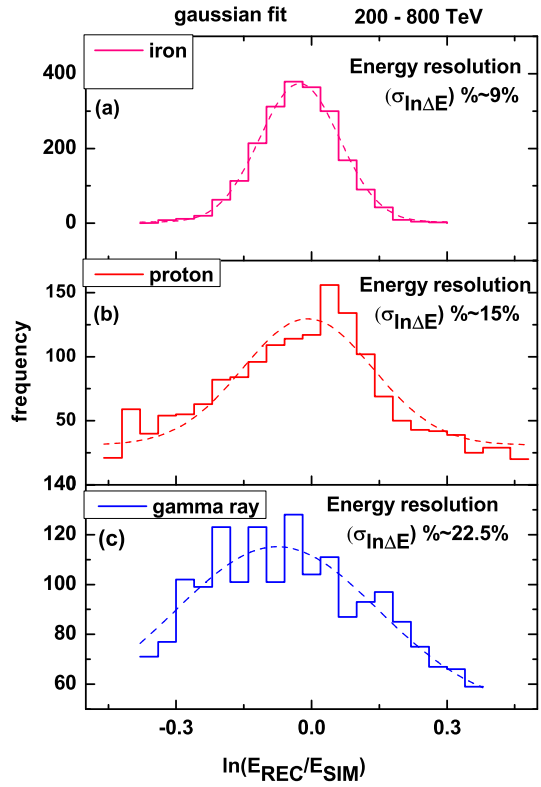} 
\end{center}
\caption{Frequency distributions of $\text{ln}(\frac{E_{\text{REC}}}{E_{\text{SIM}}})$ with $\text{E} = 200 \textendash 800$ TeV for (a) iron, (b) proton and (c) gamma ray primaries. Gaussian fits are made.} 
\end{figure}          

Finally, we evaluate performances of the relation (3) by establishing correlations of $E_{0}^{\text{REC}}$ and $E_{0}^{\text{SIM}}$ with $N_{\mu}$. Figure 5 shows the $N_{\mu}\textendash E_{0}^{\text{SIM}}$ and $N_{\mu}\textendash E_{0}^{\text{REC}}$ correlations separately for proton, iron and gamma ray initiated showers in the $E_{0}^{\text{SIM}}$-range; $200\textendash 800$ TeV. Solid blue curves in each sub-figure of the figure represent the best fitted reconstructed energy by polynomial functions. These studies suggest that the PCR energy might be estimated from the muon size and its attenuation properties in the atmosphere. It should be however mentioned that the formula (3) can also be applied to higher energy region as well except the set ($\text{a},\text{b},\text{c}$) might have taken other set of values for individual primary there. 

The energy resolution ($\sigma_{E}$) using the parameter $N_{\mu}$ is estimated at a median energy 500 TeV with $\Theta \leq 20^{\text{o}}$. The parameter $\sigma_{E}$ is obtained from the frequency distribution of $\text{ln}(\frac{E_{\text{REC}}}{E_{\text{SIM}}})$ after applying a Gaussian fit. In Figure 6, frequency distributions of $\text{ln}(\frac{E_{\text{REC}}}{E_{\text{SIM}}})$ are displayed for proton, iron and gamma ray showers in the primary energy range: $200\textendash 800$ TeV. $\sigma_{E}$ comes out as $\sim 15\%$, $\sim 9\%$ and $\sim 22.5\%$ corresponding to proton, iron and gamma ray initiated showers. Using S600 parameter the AGASA experiment yielded $\sigma_{E}$ as $30\textendash 25\%$ range at $3\times 10^7$ TeV and above. Independently using both the S500 and the combination; $N_{\text{ch}}$, $N_{\mu}$, the KASCADE group got nearly the same value for $\sigma_{E}$, which is about $22\%$. The energy resolution was also estimated using KASCADE simulation, taking $N_{\mu}$ only and the value of $\sigma_{E}$ took nearly $30\%$ at $10^{5}$ TeV. The combined data of the ARGO-YBJ and a wide field of view Cherenkov telescope using the total number of photoelectrons ($N_{\rm{pe}}$) recorded by the Cherenkov telescope yielded the energy resolution $\sim 25\%$, in their concerned energy range from 100 TeV to 3 PeV \cite{rd25}.   

\section{Conclusions}\label{S:con} 
 
In this paper, first, we have utilized some EAS observables arising out of the shower reconstruction to infer the mass composition of CRs. The reconstruction of the PCR energy by exploiting different characteristics of EASs has also been considered as the second task of the work. From the present analysis we conclude the following.

(1) The muon contents of a gamma ray shower is diminished by an average factor of $\approx 35$ relative to that expected for a hadron initiated shower. In terms of secondary hadrons the average diminution factor obtained from our simulation study is equal to $\approx 15$. This clearly suggests that muons possibly have higher discriminating power between gamma ray and hadronic primaries rather than secondary hadrons.

(2) The electron-muon and electron-hadron correlations for the KASCADE data hint for a mixed composition with a gradual transition from light to heavy mass around the \emph{knee}. A similar trend is also seen in NBU data using the electron-muon correlation alone. 

(3) The scatter plot of $\frac {\rho_{2}}{\rho_{1}}$ versus $\frac {N_{\mu}}{N_{\text {e}}}$ obtained from the multi-parameter approach is expected to be more effective to discriminate gamma ray primaries from a huge cosmic ray background.

(4) Using the estimators $N_{\mu}$ and $sec~\Theta$, it is possible to reconstruct the PCR energy. However, presence of some noticeable amount of detector inefficiency might affect the results by over- or underestimating muon sizes. Moreover, a conversion of the $N_{\mu}$ parameter of each shower into a systematic reconstructed energy is directly found from the best-fit by a polynomial function up to fourth power in $N_{\mu}$.

(5) The energy resolution for the derived energy ($E \geq 200$ TeV) using the muon size is a promising result and might be a nice starting point for possible future improvements from the point of view of its application.   
 
\section*{Acknowledgments}
 RKD and AB thank the SERB, Department of Science and Technology (Govt. of India) for financial support under the Grant no. EMR/2015/001390.


\begin{thebibliography}{99}

\bibitem{rd1} T Antoni \emph{et al.} \emph{Nucl. Instrum. Methods A} \textbf{513} 490 (2003).
\bibitem{rd2} S K Gupta \emph{et al.} \emph{Nucl. Instrum. Methods A} \textbf{540} 311 (2005).
\bibitem{rd3} G Disciascio and T Di Girolamo \emph{Astrophys. Space Sci.} \textbf{309} 537 (2007).
\bibitem{rd4} T Antoni \emph{et al.} \emph{Astropart. Phys.} \textbf{24} 1 (2005).
\bibitem{rd5} K Greisen \emph{Prog. in Cosmic Ray Physics, North Holland, Co. Amsterdam} \textbf{3} 1 (1956).
\bibitem{rd6} P Blasi \emph{Nucl. Phys. B, Proc. Suppl.} \textbf{239} 140 (2013).
\bibitem{rd7} R Aloisio, V Berezinsky and A Gazizov \emph{Astropart. Phys.} \textbf{39} 129 (2012).
\bibitem{rd8} W D Apel \emph{et al.} \emph{Phys. Rev. Lett.} \textbf{107.171104} (2011).
\bibitem{rd9} A Bhadra, S K Sarkar, C Chakraborty, B Ghosh and N Chuodhuri \emph{Nucl. Instrum. Methods A} \textbf{414} 233 (1998).
\bibitem{rd10} N N Kalmykov, S S Ostapchenko and A I Pavlov \emph{Nucl. Phys. B (Proc. Suppl.)} \textbf{52} 17 (1997)
\bibitem{rd11} K Werner \emph{et al.}: \emph{Phys. Rev. C}, \textbf{74} 044902 (2006).
\bibitem{rd12} H Fesefeldt \emph{Report PITHA-85/02 (RWTH Aachen)} (1985).
\bibitem{rd13} D Heck, J Knapp, J N Capdevielle, G Schatz and T Thouw, FZKA report-6019 ed. FZK \emph{The CORSIKA  Air Shower Simulation Program} Karlsruhe (1998).
\bibitem{rd14} W R Nelson, H Hiramaya, D W O Rogers \emph{Report SLAC 265} (1985).
\bibitem{rd15}  National Aeronautics and Space Administration (NASA) US Standard Atmosphere Technical Report \textbf{NASA - TM - X - 74335} (1976).
\bibitem{rd16} R K Dey, A Bhadra and J N Capdevielle \emph{J. Phys. G: Nucl. Part. Phys.} \textbf{39} 085201 (2012). 
\bibitem{rd17} A Bhadra \emph{Pramana: J. Phys.} \textbf{52} 2 (1999).
\bibitem{rd18} S Lafebre, R Engel, H Falcke, J Hoerandel, T Huege, J Kuijpers and R Ulrich, \emph{Astropart. Phys.} \textbf{31} 243 (2009).
\bibitem{rd19} T Cheung and P K Mackeown, \emph{Nuovo Cim. C} \textbf{11} 193 (1988).
\bibitem{rd20} T Stanev, T K Gaisser and F Halzen, \emph{Phys. Rev. D} 32 1244 (1985).
\bibitem{rd21} T V Danilova \emph{et al.} \emph{Proc. 19th Int. Cos. Ray Conf.} \textbf{2} 260 (1985).
\bibitem{rd22} J Alvarez-Muniz \emph{et al.} \emph{Phys. Rev. D} \textbf{69} 103003 (2004).
\bibitem{rd23} T Pierog and K Werner \emph{Phys. Rev. Lett.}, \textbf{101} 171101 (2008).
\bibitem{rd24} T Antoni \emph{et al.} \emph{Astropart. Phys.} \textbf{19} 703 (2003).
\bibitem{rd25} B Bartoli \emph{et al.} \emph{Phys. Rev. D} \textbf{92} 092005 (2015).
\bibitem{rd26} R K Dey and S Dam \emph{Eur. Phys. J. Plus} \textbf{131} 402 (2016).
\bibitem{rd27} A W Wolfendale {\it Nucl. Phys. B Proc. Suppl.} {\bf 22 80} (2015).
\bibitem{rd28} S Vernetto and P Lipari \emph{Phys. Rev. D} \textbf{94} 063009 (2016).
\end{thebibliography}
\end{document}